\documentclass[aps,pre,twocolumn,superscriptaddress,floatfix,showpacs]{revtex4}

\bibstyle{apsrev.bib}

\usepackage{amssymb}
\usepackage{amsmath}
\usepackage{latexsym}
\usepackage{graphicx}
\usepackage{subfigure}

\begin{document}

\title{Elastic lines on splayed columnar defects studied numerically}

\author{Viljo Pet\"aj\"a}
\affiliation{
Helsinki University of Techn., Lab. of Physics,
P.O.Box 1100, 02015 HUT, Finland}
\author{Matti Sarjala}
\affiliation{
Helsinki University of Techn., Lab. of Physics,
P.O.Box 1100, 02015 HUT, Finland}
\author{Mikko Alava}
\affiliation{
Helsinki University of Techn., Lab. of Physics,
P.O.Box 1100, 02015 HUT, Finland}
\author{Heiko Rieger}
\affiliation{
Theoretische Physik, Universit\"at des Saarlandes, 
66041 Saarbr\"ucken, Germany}

\begin{abstract}
We investigate by exact optimization method properties of two- and
three-dimensional systems of elastic lines in presence of splayed
columnar disorder. The ground state of many lines is separable both in
2d and 3d leading to a random walk -like roughening in 2d and
ballistic behavior in 3d. Furthermore, we find that in the case of
pure splayed columnar disorder in contrast to point disorder there is
no entanglement transition in 3d. Entanglement can be triggered by
perturbing the pure splay system with point defects.
\end{abstract}

\pacs{74.25.Qt, 05.40.-a, 74.62.Dh}

\maketitle

\newcommand{\bc}{\begin{center}}
\newcommand{\ec}{\end{center}}
\newcommand{\be}{\begin{equation}}

\newcommand{\ee}{\end{equation}}
\newcommand{\ba}{\begin{array}}
\newcommand{\ea}{\end{array}}
\newcommand{\beqn}{\begin{eqnarray}}
\newcommand{\eeqn}{\end{eqnarray}}

\section{Introduction}

Transport properties of type-II superconductors are influenced
by the presence of various kinds of disorder \cite{Blatter}.
Pinning of vortex lines hinders their motion, which,
in response to an applied current, causes dissipation.
From the practical point of view it is highly desirable to avoid the
appearance of vortex creep which gives rise to a finite
resistivity. It was proposed by Hwa {\it et al.} \cite{Hwa} that 
splayed columnar defects resulting from heavy ion
irradiation of superconducting samples, would significantly enhance 
the vortex pinning, and thus reduce the vortex creep leading to a higher
critical current density $j_c$.
An additional important aspect might be the inhibition of
vortex motion due to forced entanglement induced 
by the disorder \cite{entgl}. 

The predictions concerning $j_c$ have been been verified in
experiments on samples with different
sources of splayed defects \cite{K-E,Civale,Kwok}.
Kwok et al. \cite{Kwok} reported that well above the matching
field $B_{\phi}$, where the density of vortices equals to the density
of columnar defects, the irreversibility line of sample with splay is
below of the one without splay defects. As an evidence of strong
entanglement above the matching field it was discovered \cite{Kwok}
that although the irreversibility is decreased in samples with splay
defects values of $j_c$ are still greatly increased compared to
unirradiated samples.  On the other hand the comparison of samples
with one and two families of splay defects did not reveal any
differences in the values of $j_c$
\cite{Hebert}.
Molecular dynamics simulations for $B>B_{\phi}$ show the increase of
$j_c$ in the samples with splay disorder
\cite{Palmer,Olson}. The authors of \cite{Olson} have also performed
simulations on samples with $B<B_{\phi}$ for which they obtained a
milder enhancement of $j_c$. On 
the basis of these observations they suggested that in 
the case $B>B_{\phi}$ the additional increase of $j_c$ is due to
vortex entanglement.

The ground state properties of an {\em ensemble} of flux lines
in such disordered environments has never been investigated before.
Single flux line properties in the presence of
tilted columnar defects at zero temperature
were studied by Lidmar et al. \cite{Lidmar}. They show that the behavior 
of the lines depends on the energy distribution of the lines.
This is manifested in roughening, or mean-square displacement
as a function of sample height $H$,
\begin{equation}
w^2(H) = \langle \mathrm{\bf
r}(z)^2\rangle_z-\langle \mathrm{\bf r}(z)\rangle_z^2 \sim H^{2\zeta}.
\end{equation} 
Here $z$ is the distance along the height direction, and $r$ the
transverse displacement. According to Lidmar et al. the roughness
exponent $\zeta$ is sensitive to the shape of the distribution of the 
tilt angle and the energy distributions of the defects \cite{Lidmar}.  
For instance with an opening angle of $90^{\circ}$ and for a uniform
energy distribution the roughness exponent in 2d is $\zeta=3/4$, in
contrast to the point-disorder result $2/3$ \cite{Halpin}.
In samples with a fixed starting point a single line has the 
following geometry. It occupies a splay defect until a jump to a 
energetically more favorable one takes place. 
The lines undergo jumps from splay defect to splay defect so that the
average distance between two successive jumps grows as $\Delta z \sim
z$ \cite{Lidmar}. Thus though the jump density decreases with growing
$z$ the  roughening exhibits a non-trivial scaling.  

The natural question arises how the single-line physics
outlined above changes in the presence of many interacting
lines, at a constant line density $\rho$. 
In this paper we study this in both two (2d) and three dimensions (3d). 
There
are studies of the role of disorder in two-dimensional
samples \cite{Bolle}, while the three-dimensional case corresponds
to bulk superconductors. We find that at finite line densities 
the physics changes in particular as the roughening of lines
is concerned: the roughness exponent becomes $\zeta=1/2$
in 2d and $\zeta=1$ in 3d. The ballistic behavior of the lines leads
to the absence of true entanglement in 3d. 

This paper is organized as follows.
In Sec.\ \ref{sec:model} we explain the
details of our numerical model. The scaling of roughness is discussed
in Sec.\ref{sec:rough}. We also study in Sec. \ref{sec:sep} if
inserting a new line causes rearrangements in the
configuration of previously inserted lines, ie. as the 
density is increased.
 The results shown in Sec.\ \ref{sec:entgl} demonstrate
 that in pure splay disorder lines do not
entangle. Entanglement can be induced by perturbing splay with point
disorder. The
summary and discussion are presented in Section \ref{sec:summ}.

\section{Model}\label{sec:model}

\begin{figure}[t]\centering
  \hfill
  \includegraphics[width=0.42\columnwidth]{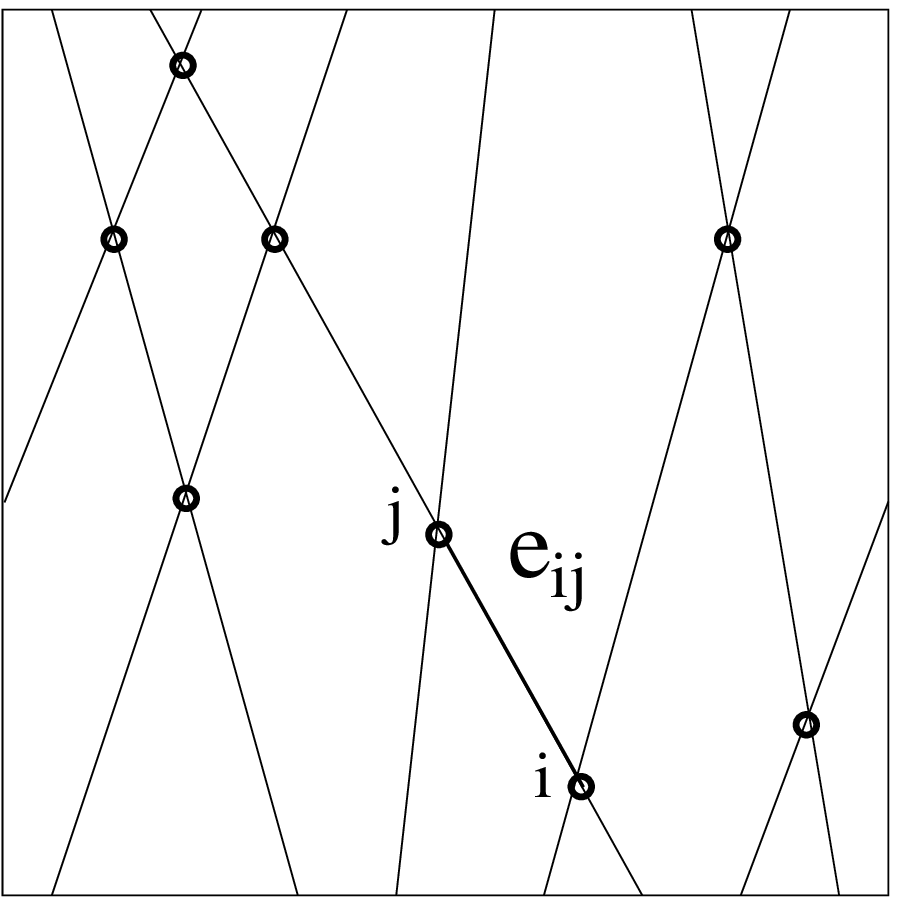}
  \hfill
  \includegraphics[width=0.42\columnwidth]{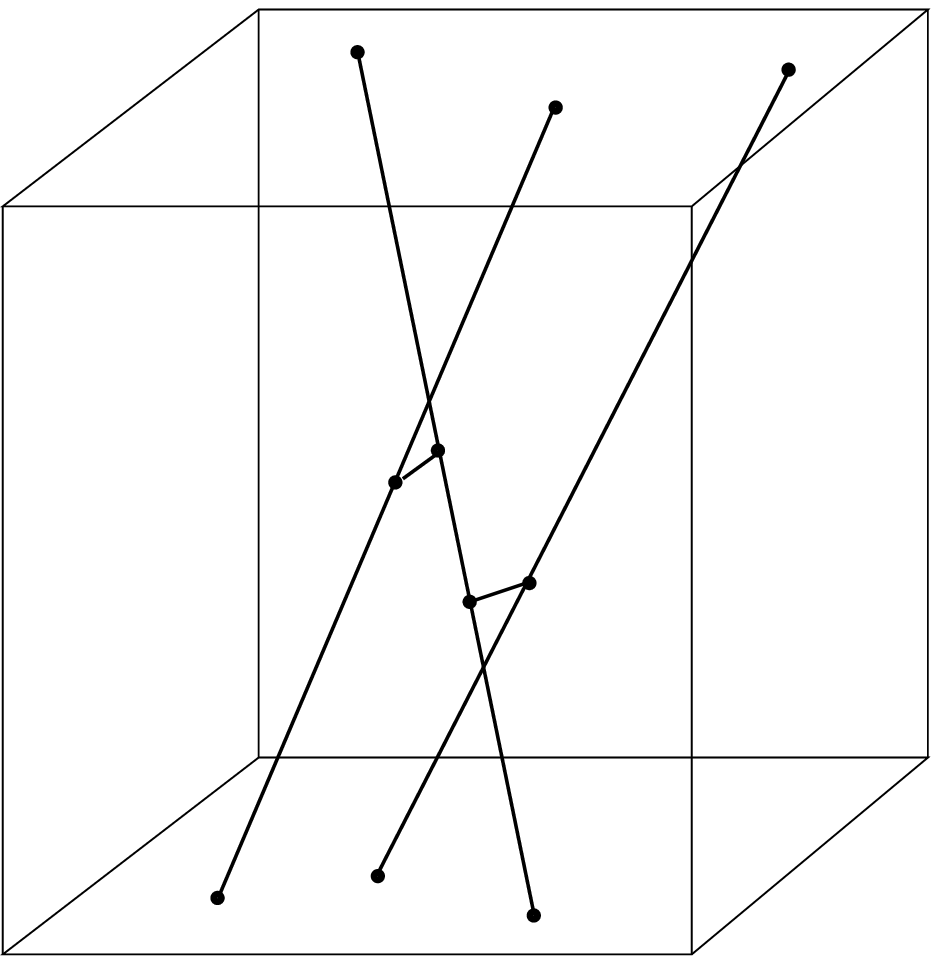}
  \hfill
  \caption{\label{graph} Schematic figure of the graph consisting of
    splayed columns in two and three dimensions. In the latter case
    note the two additional bonds connecting the columns.}
\end{figure}

We consider the following model of a system of interacting lines in a
two- or three-dimensional disordered environment: The lines live on
the bonds of a graph consisting of an ensemble of splayed columns
embedded in a box with a width $L$ and a height $H$
(Fig.\ \ref{graph}). Each column is described with a transverse
coordinate $\mathrm{\bf r}(z)$ at height $z$ from the bottom level:
\be 
\mathrm{\bf r}(z)=\mathrm{\bf r}_0 + \mathrm{\bf a}\, z 
\ee 
where $\mathrm{\bf r}_0$ is a random point on the basal plane and
$\mathrm{\bf a}$ is a randomly chosen variable that defines the
magnitude and the orientation of the tilt of a given column.  

The number of columnar defects $M$ is set to $M=L$ and $M=L^2$ in two and 
three dimensions respectively. The average distance
between two nearby columnar defects defines the in-plane length unit
of the model. The lines enter the system at the bottom plane ($z=0$)
and they exit at the top plane ($z=H$), and we use both fixed and free
entrance points, and free exit points. The case of a fixed entry point
and a free exit point is that considered by Lidmar et al. for a single
line \cite{Lidmar}, who discuss an experimental scenario for the same.
Also rough enough surface could pin the end points of flux lines. 
Strong pinning along large surface steps has been reported in 
Ref.\ \cite{dai}. More recent experimental results \cite{fil} indicate
that surface pinning have measurable effects on the flux lattice
dynamics.

In this
paper we study the case where the columns have a random, uniformly
distributed orientation within a cone with a fixed opening angle. In
our model changing the opening angle of the cone is equivalent to a
rescaling of the system height, for which reason we consider only the
opening angle of $90^{\circ}$. 

The number $N$ of lines threading the sample is fixed by
the prescribed density $\rho=N/M$. Within this version
of the model number of lines cannot exceed the number of columns, which
means in the case of splay disorder that $B<B_{\phi}$.
The graph could be in principle modified such that lines can traverse
the system also  between the columnar defects which would correspond
to $B>B_{\phi}$. 

In the transverse directions we use both periodic and open boundary
conditions. In the latter case the defects crossing the boundary are cut
such that lines cannot follow them across the system
boundaries. Note that one can not let flux lines escape from truly
open boundaries since the line density would decrease with $z$,
the longitudinal coordinate. In order to avoid correlations in systems 
with $H>L$ we first construct a graph of size $H^d$ from which we cut a 
piece of size $L$.

In 2d lines can change defects only at their crossings and in 3d when
the defects are close enough to each other.  In 3d columns are
connected by introducing an extra segment between the lines at the
shortest mutual distance whenever this distance is shorter than a
fixed value $r_c$.  When $r_c$ is kept relatively small compared to
the average distance between the columnar defects the choice of the
energy cost of the bond connecting two defects can be arbitrary. We
choose zero energy cost and $r_c=1.4$. We have also checked the value
$r_c=0.5$ which did not change the results.

We model the disordered environment by assigning a random (potential)
energy $e_{ij}=r_{ij} u$ to each bond $(i,j)$, where $r_{ij}$ is the
Euclidean length of the bond and $u$ is a random number which
determines the type of disorder.  In the case of splay disorder the
random variable $u$ is drawn independently for each column such that
bonds along a given column share the same value of $u$. Point disorder
is modeled with a similarly constructed graph. The only difference is
that $u$ is drawn independently for each bond which destroys the
correlations along the columns. For point disorder we use uniformly
distributed $u$ and for splay disorder the following distribution
in order to make comparisons to single line results by Lidmar et al.
\cite{Lidmar}:
\be
P(u)=\nu u^{\nu-1} \,, 
\label{eq:distr}
\ee
for $0<u<1$ and otherwise $P(u)=0$. With $\nu=1$ this reduces simply
to a uniform distribution. 

For computational convenience we focus on the short screening
length limit \cite{screening}.
We restrict ourselves to hard-core interactions between the lines,
which means that their configuration is specified by bond-variables
$x_{ij}\in\{0,1\}$, $x_{ij}=1$ indicating that a line segment occupies
a bond between nodes $i$ and $j$, and $x_{ij}=0$ indicating that no
line segment occupies this bond. 
The total energy of the line configuration is given by
\be
{\mathcal H}=\sum_{(ij)} x_{ij} \cdot e_{ij} \, ,
\label{ham}
\ee
where the summation is performed over all bonds. The corresponding
Hamiltonian in a continuum limit is given by the following formula:
\beqn
{\mathcal H} = {\displaystyle\sum_{i=1}^N}
{\displaystyle \int_0^H dz }
\Bigl\{ \frac{\gamma}{2}\left[\frac{d{\bf r}_i}{dz}\right]^2
+V_r[{\bf r}_i(z),z] \nonumber \\
+\sum_{j(\neq i)}V_{\rm int}[{\bf r}_i(z)-{\bf r}_j(z)]
\Bigr\}\; ,
\label{cont}
\eeqn
where ${\bf r}_i(z)$ denotes the transversal coordinate at
longitudinal height $z$ of the $i$-th flux line. The interactions
$V_{\rm int}[{\bf r}_i(z)-{\bf r}_j(z)]$ are hard core repulsive and
the disorder potential $V_r[{\bf r}_i(z),z]$ is $\delta$-correlated in
the case of point disorder and strongly correlated along columns in
the case of columnar disorder. The elastic energy is mimicked in our
numerical model by the positivity of all energy costs per unit length,
which ensures for instance the minimization of the line length in the
absence of randomness.

\begin{figure}[t]\centering
  \includegraphics[width=0.9\columnwidth]{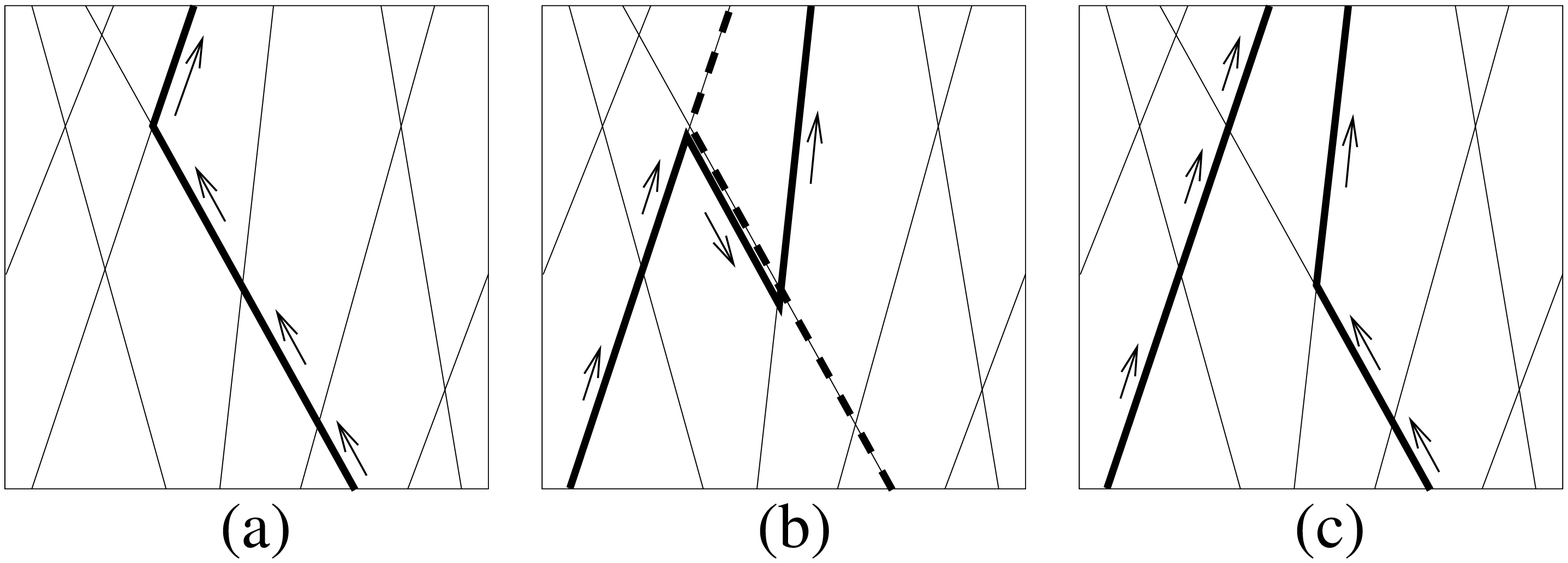}
  \caption{\label{flow} Schematic figure of adding lines to the graph.
    (a) The ground state of a single line with arrows indicating the direction
    of the optimal path. (b) The bold line is the optimal path from
    bottom to top in the residual graph. The residual graph is 
    constructed by changing the energy costs of the bonds to negative
    along the dotted line. In addition, the direction of paths in the
    residual graph on the dotted bonds must be opposite to the
    direction of the previously inserted path shown in (a). 
    (c) The final line configuration after the path segments with
    opposite directions are canceled.} 
\end{figure}

At low temperatures the line configurations will be dominated by the
disorder and thermal fluctuations are negligible. Therefore we
restrict ourselves to zero temperature and focus on the ground state
of the Hamiltonian (\ref{ham}). Computing the ground state now
corresponds to finding $N$ non-overlapping directed paths traversing
the graph along the bonds from bottom to top. One has to minimize the
total energy of the whole set of the paths and not of each path
individually (n.b.: already the two-line problem is actually
non-separable \cite{twoline}). For a single line problem the lowest
energy path can be found straightforwardly with Dijkstra's shortest
path algorithm. In the case of many lines Dijkstra's algorithm is
applied successively on a residual graph instead of the original graph
\cite{opt-review} as illustrated in Fig.\ \ref{flow}.  In the residual
graph the properties of occupied bonds are modified as follows: If
$x_{ij}=1$ we set $e_{ij}^{\rm residual}=-e_{ij}$, i.e. negative, and
require that the direction of a new path must be in the opposite
direction with respect to the direction of previously inserted
paths. The path segments with two opposite directions on a given bond
are canceled. For details see \cite{opt-review}.

Although the lines cannot occupy the same bond of the lattice they may
touch in isolated points if the nodes in these points have 
more than three neighbors.
This is now the case only in 2d as exemplified in
Fig.\ref{graph}. Since we want to calculate the roughness of lines we need to
determine the individual lines, for which we use a local rule. In 2d
the line identification is unambiguous if we simply require that the
lines cannot cross.  

In our model flux lines are confined to the defects with no
possibility to enter the bulk (except in 3d if the jumps from one
defect to another are counted as such). We also tested the case
where a homogenous bulk between the splayed defects is represented by a
set of densely packed columnar defects with zero tilt and with a
constant energy cost per unit length. Here, after a flux line has found a
splayed defect with a lower energy cost per unit length in
$z$-direction compared to the one in the bulk one
recovers the pure splay disorder behavior.  Thus, including the
possibility for lines to travel also in the bulk introduces another
crossover length. We have checked numerically that this scenario
holds. After a cross-over system size, which depends on the ratio of
the energy costs in the bulk and on the defects, for single line
systems we obtain the roughness exponents of pure splay disorder.

In many line systems with a fixed ratio between the number of splay
defects and the number of inserted flux lines it is not guaranteed
that all lines can find a defect which is energetically more
favorable than the bulk. Depending on the energy cost of the bulk
there can be only a fraction of splayed defects which are
energetically more favorable compared to the bulk. If the number of
such defects is smaller than the number of flux lines there will be a
fraction of lines $f_{bulk}$ that stay in the bulk throughout the
whole sample. As a function of the energy cost per unit length in the
bulk $u_{bulk}$ this scales like $f_{bulk}=g(u_{bulk}/\rho^{\alpha})$
with $g(x)$ a scaling function that decreases monotonically
from $1$ at $x=0$ to zero for $x=1$ and $\alpha$ depending on the shape of
the energy distribution of the splayed defects (Eq.(\ref{eq:distr})).

Since the lines staying in the homogenous bulk have no transverse
fluctuations our central results on the geometrical properties of flux
line systems in splay disorder are not expected to change qualitatively.

\section{Roughness of the ground state}\label{sec:rough}

In this section we focus on the scaling of the average roughness $w$,
i.e. the amount of transverse wandering of the lines. The mean square
displacement the $i$-th line is $w^2_i=\langle \mathrm{\bf
r}_i(z)^2\rangle_z-\langle \mathrm{\bf r}_i(z)\rangle_z^2$ where
$\mathrm{\bf r}_i(z)$ is the transverse coordinate of the line $i$ at the
distance $z$ from the bottom level and
$\langle\mathrm{\bf r}(z)\rangle_z=\frac{1}{H}\int_0^H \mathrm{\bf
  r}(z)\, dz\,$ denotes the average 
along the line from the bottom $z=0$ to the top $z=H$. We define $w^2$
as the disorder average of $w^2_i$ averaged over all lines.

In the case of point disorder our model is in the same universality
class as the directed polymer model according to which the roughness of
one line scales as $w\sim H^{\zeta}$ with well known exponents
$\zeta_{2d}=2/3$ and $\zeta_{3d}\approx 5/8$ in two and three
dimensions respectively \cite{Halpin}. 
In 2d the
steric repulsion between the lines leads to 
collective rearrangement of the lines which yields a
logarithmic growth of the roughness. In 3d lines can
wind around each other which suppresses the repulsion resulting in
a random walk-like behavior of lines \cite{rough}. 

For splay disorder, we propose the following simple scenario.
At small $z$ lines do not see each other and exhibit single
line behavior. Beyond some value of $z$,
which depends on the density $\rho$,
the lines cannot further optimize their
configurations and stay on the same defects.
This leads to a linear growth of the roughness, $\zeta =1$ in
3d due to ballistic behavior. 
The same arguments on the structure of the optimal line configuration 
hold also for 2d. However, in 2d splayed defects can cross each other
while the individual flux lines are identified such that they
do not cross the other flux lines.
Thus, in 2d one has effectively system of
hard-core repulsive random walks resulting in $\zeta=1/2$.
Thus one would expect a roughness scaling form 
\begin{equation}
w(L,H) \propto L^\zeta f(H/L),
\label{eq:scaling}
\end{equation}
where $f(x)$ is a scaling function. In both 2d and 3d,
this scenario is independent of the splay energy distribution.

\begin{figure}[t]\centering
\includegraphics[angle=-90,width=0.99\columnwidth]{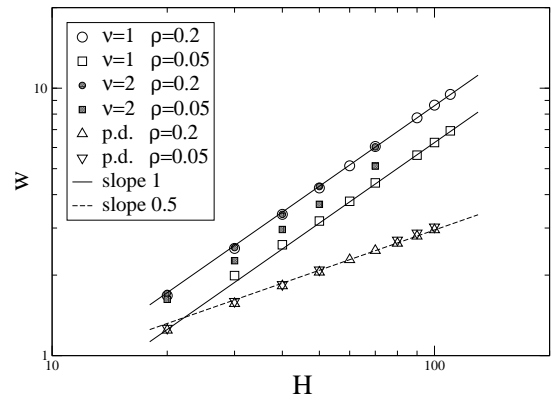}
\includegraphics[angle=-90,width=0.99\columnwidth]{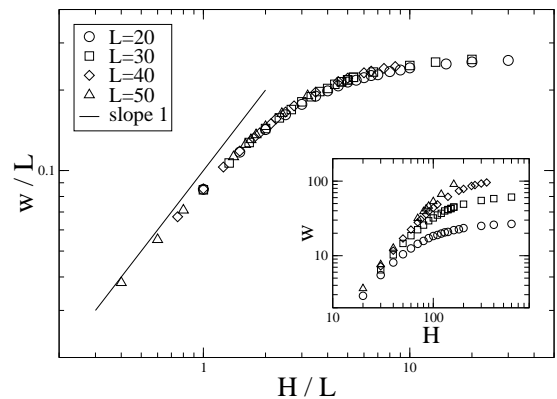}
\caption{\label{r3d} 
  Doubly logarithmic
  plots for the roughness $w$ of multi-line systems in 3d.
  Top:
  Roughness vs. system height $H$ ($L=H$) for different values of line
  density $\rho$ and for two defect energy distributions ($\nu=1,\,
  2$); lines have fixed starting points. For comparison are shown also
  the roughness for point disorder (labeled as ``p.d.''). 
  Bottom: Scaling plot for the
  roughness with fixed system widths for $\rho=0.2$ and $\nu=1$, the
  un-scaled data are shown in the inset. At this line density the
  difference between the data with fixed and free starting points is
  undistinguishable.} 
\end{figure}

Fig.\ \ref{r3d} shows the correctness of this proposition in 3d.
One can see from Fig.\ \ref{r3d}(a) that the roughness of many line
systems grows linearly $w\sim H$ with no dependence on
the energy distribution of the defects.
According to the data collapse in Fig.\ \ref{r3d}(b) the
saturation roughness and the saturation height grow linearly
with the system width $L$, cf. with Eq.(\ref{eq:scaling}).

\begin{figure}[t]\centering
\includegraphics[angle=-90,width=0.99\columnwidth]{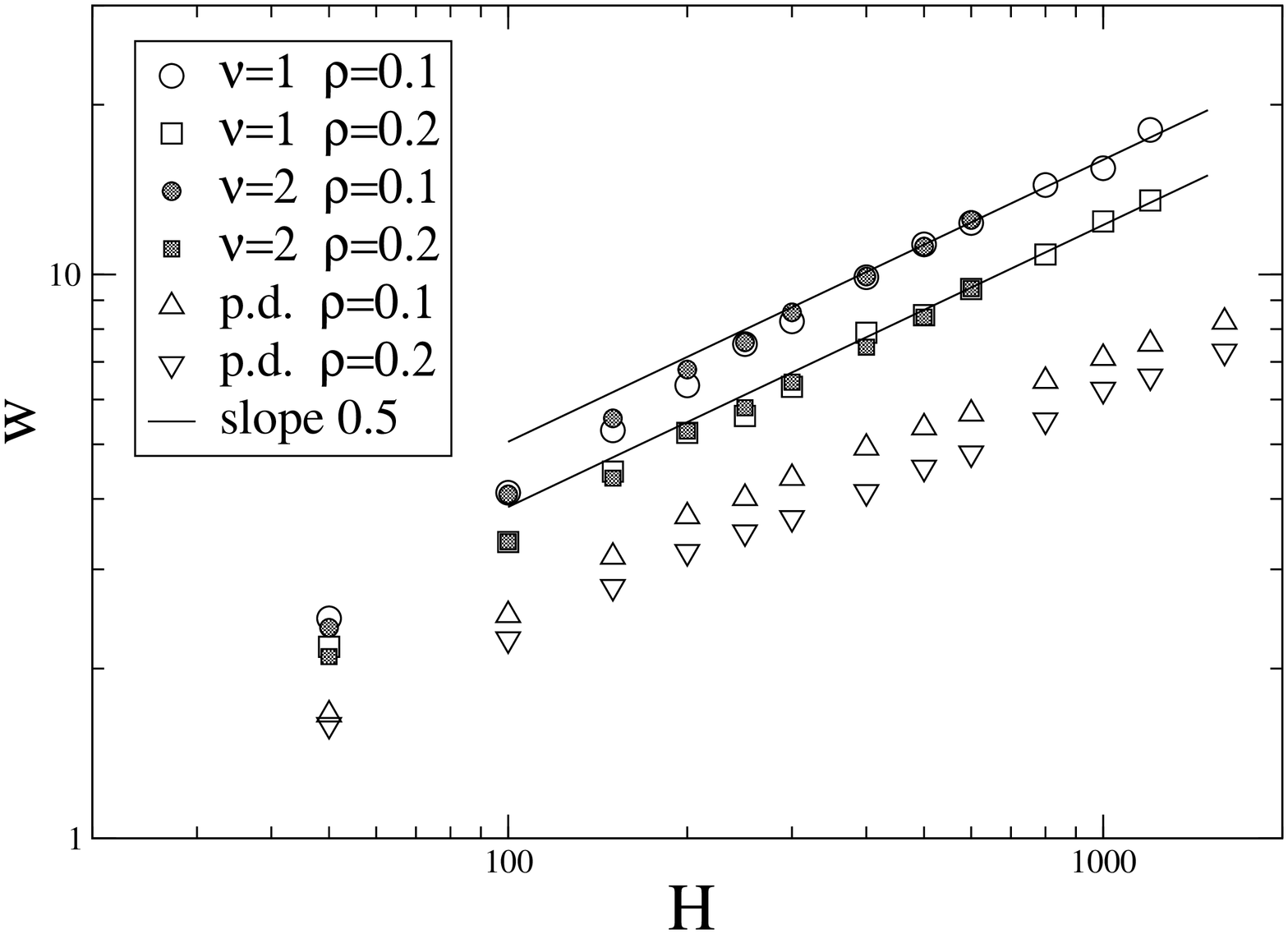}
\includegraphics[angle=-90,width=0.99\columnwidth]{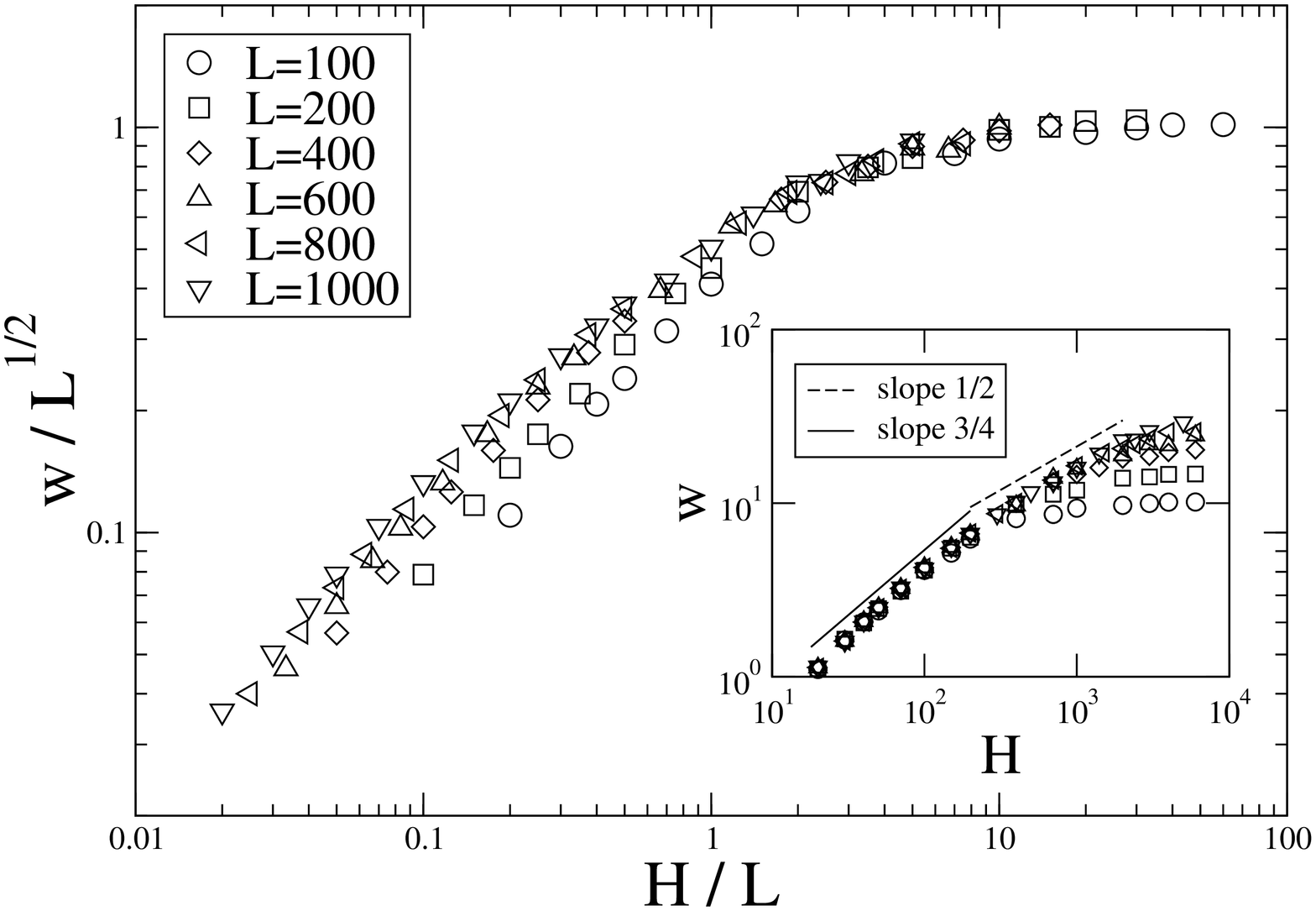}
\caption{\label{r2d} 
  Double logarithmic plots for the roughness $w$ of multi-line 
  systems in 2d.  
  Top:
  Roughness vs. system height $H$ ($L=H$) for different values of line
  density $\rho$ and for two defect energy distributions ($\nu=1,
  2$); lines have fixed starting points. 
  Bottom: Scaling plot of the roughness with fixed system
  widths $L$ and fixed starting points with line density $\rho=0.1$
  and $\nu=1$.} 
\end{figure}

In 2d there is a collective regime,
where the lines exhibit random walk like -behavior as suggested
above.
Fig.\ \ref{r2d}(a) shows that the roughness grows asymptotically
like $w\sim H^{0.5}$ for both values of $\nu$.
The data collapse shown in Fig.\ \ref{r2d}(b) gives the random walk
like scaling also for the saturation roughness in agreement with Fig.\
\ref{r2d}(a). 
Close to the system boundaries the roughness of the lines is
suppressed (due to the way the line identification is made in 2d)
which shows up in the scaling of small system widths $L$.

\section{Separability of the ground state}\label{sec:sep}
We define that the ground state configuration of
$N$ lines is separable if it can be obtained by adding successively
flux lines to the system without modifying the configurations of
the previous lines. This can be checked from the successive
shortest path algorithm: whenever
segments of flux lines are canceled in the residual
graph this implies that adding a line changed the previous
configuration (see Fig.\ \ref{flow}). Hence we focus on such
segments or bonds and 
calculate the sum of the energy costs $E_s=\sum e_{ij}$ of such bonds
and use it as a measure of the separability. When $E_s=0$ no flux is
canceled due to possible rearrangements of line configuration. This
means that one has fully separable ground state.

In the case of (splayed) columnar disorder it is obvious that with
periodic boundary conditions and the full freedom of choosing the most
favorable starting point the lines pick up the defects in the order of
their total energy cost from the source to the target. Hence, no
rearrangements of flux line configurations is needed. Introducing the
boundaries or any other distortions to the pure columnar disorder
reduces the separability of the ground state.

\begin{figure}[t]\centering
\includegraphics[angle=-90,width=0.99\columnwidth]{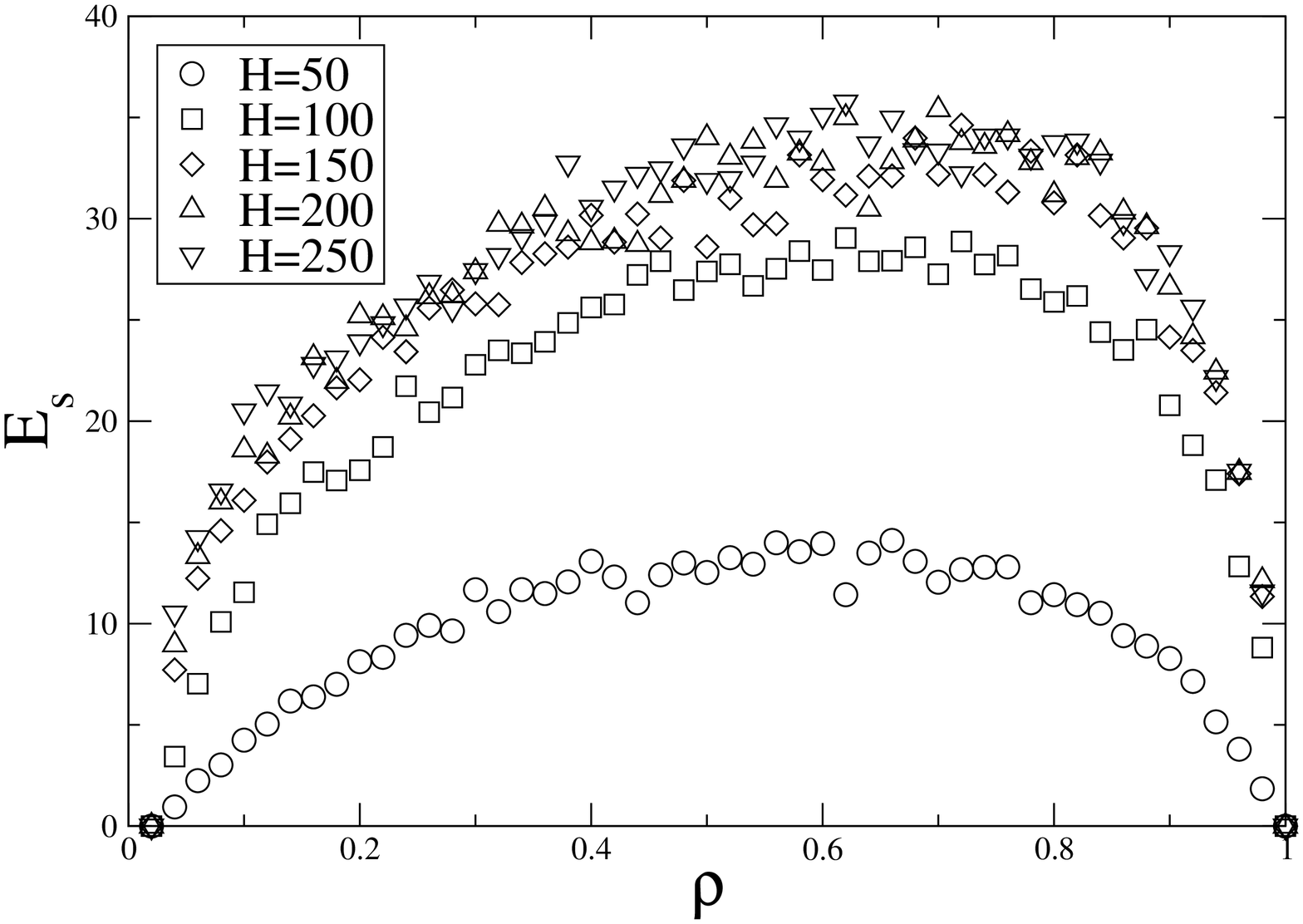}
\includegraphics[angle=-90,width=0.99\columnwidth]{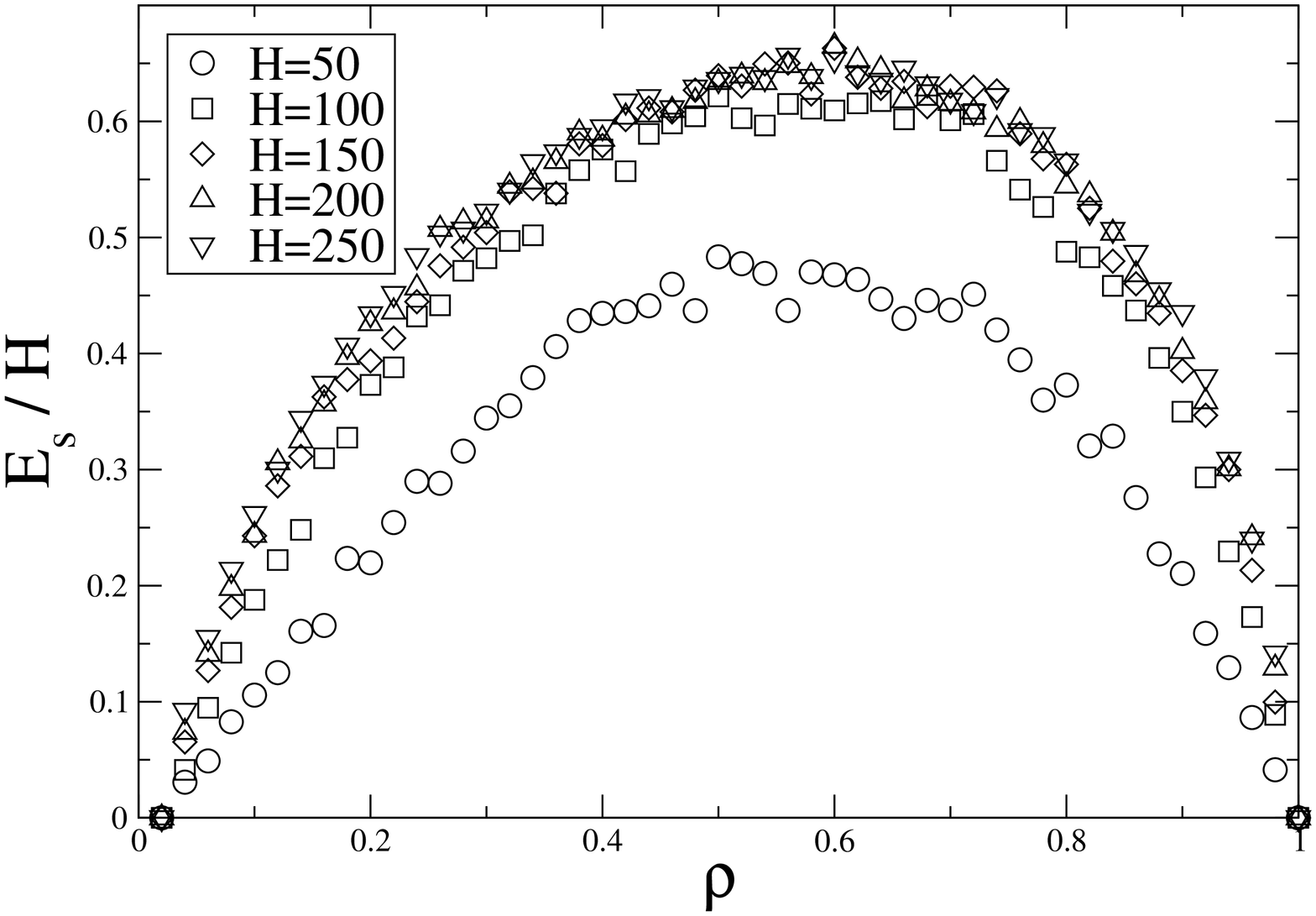}
\caption{\label{sep2d} A measure of the separability of the ground state in 2d,
  periodic boundaries, fixed starting points, L=50. Top: splay
  disorder, $E_s$ saturates as $H$ grows. Bottom: point disorder,
  $E_s$ grows linearly with $H$.}
\end{figure}

\begin{figure}[t]\centering
\includegraphics[angle=-90,width=0.99\columnwidth]{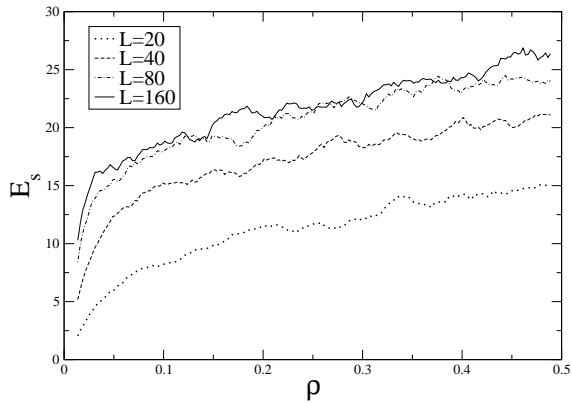}
\includegraphics[angle=-90,width=0.99\columnwidth]{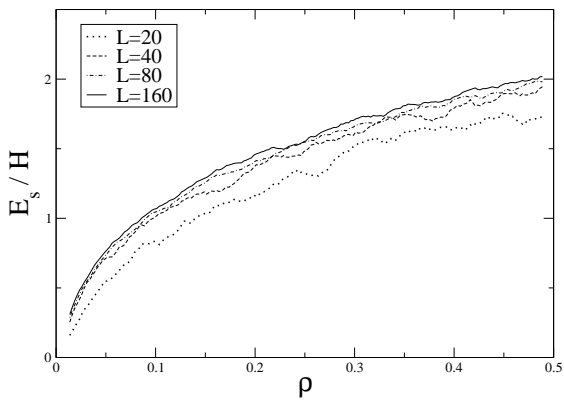}
\caption{\label{sep3d} Same as Fig.\ \ref{sep2d}, now in 3d with
  L=20 and line density only up to $\rho=0.5$ due to increased
  computational costs. The data is smoothened  by taking running 
  averages of each 10 subsequent data points.}
\end{figure}

In Fig.\ \ref{sep2d} we demonstrate the difference between the
separability of the ground states of splay and point disorder.  We
consider $L\times H$ systems in two dimensions with periodic boundary
conditions in the direction perpendicular to the $z$-axis.  In the
case of splay disorder $E_s$ saturates at a particular system height,
because line rearrangements do not take place at greater $z$. Thus
$E_s$ divided by the system volume goes to zero with increasing
$H$, implying separability.
Compare with the fact that $E_s$ is linear in
$H$ in the case of point disorder, indicating non-separability.
We calculated $E_s$
also in three dimensions and observed the same
behavior as in the 2d case as is depicted in Fig. \ref{sep3d}.

\section{Entanglement}\label{sec:entgl}

\begin{figure}
  \begin{center}
    \mbox{
      \subfigure[]{
	\includegraphics[width=0.325\columnwidth]{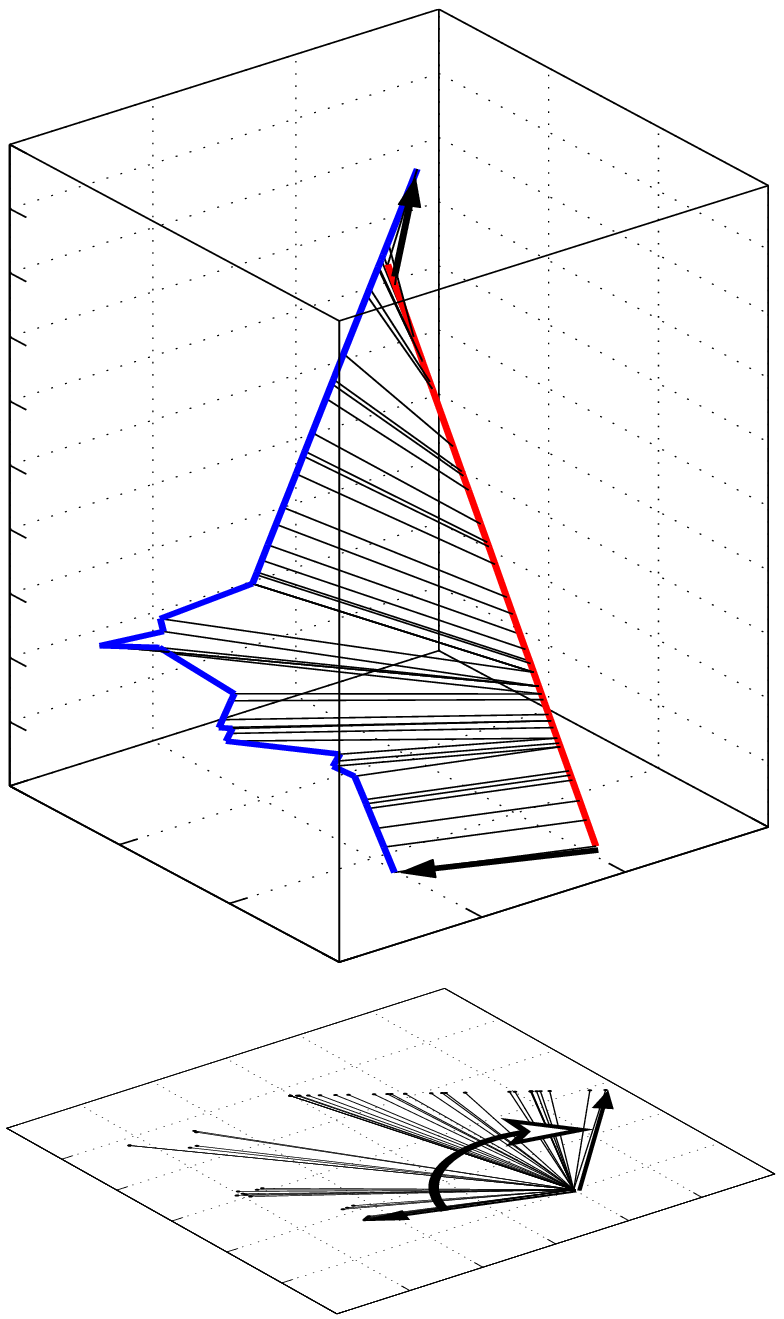}
      }
      \hspace{-0.41cm}
      \subfigure[]{
	\includegraphics[width=0.325\columnwidth]{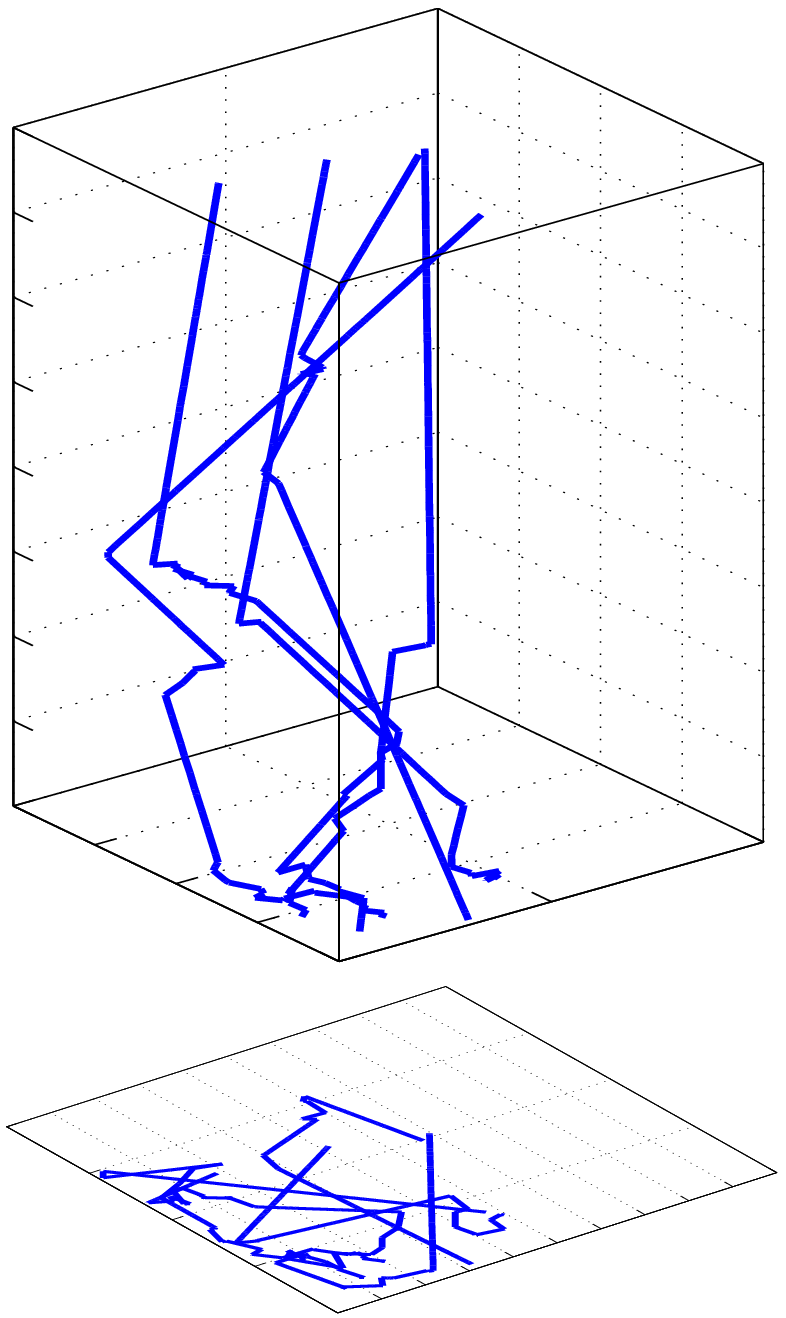}
      }
      \hspace{-0.42cm}
      \subfigure[]{
	\includegraphics[width=0.325\columnwidth]{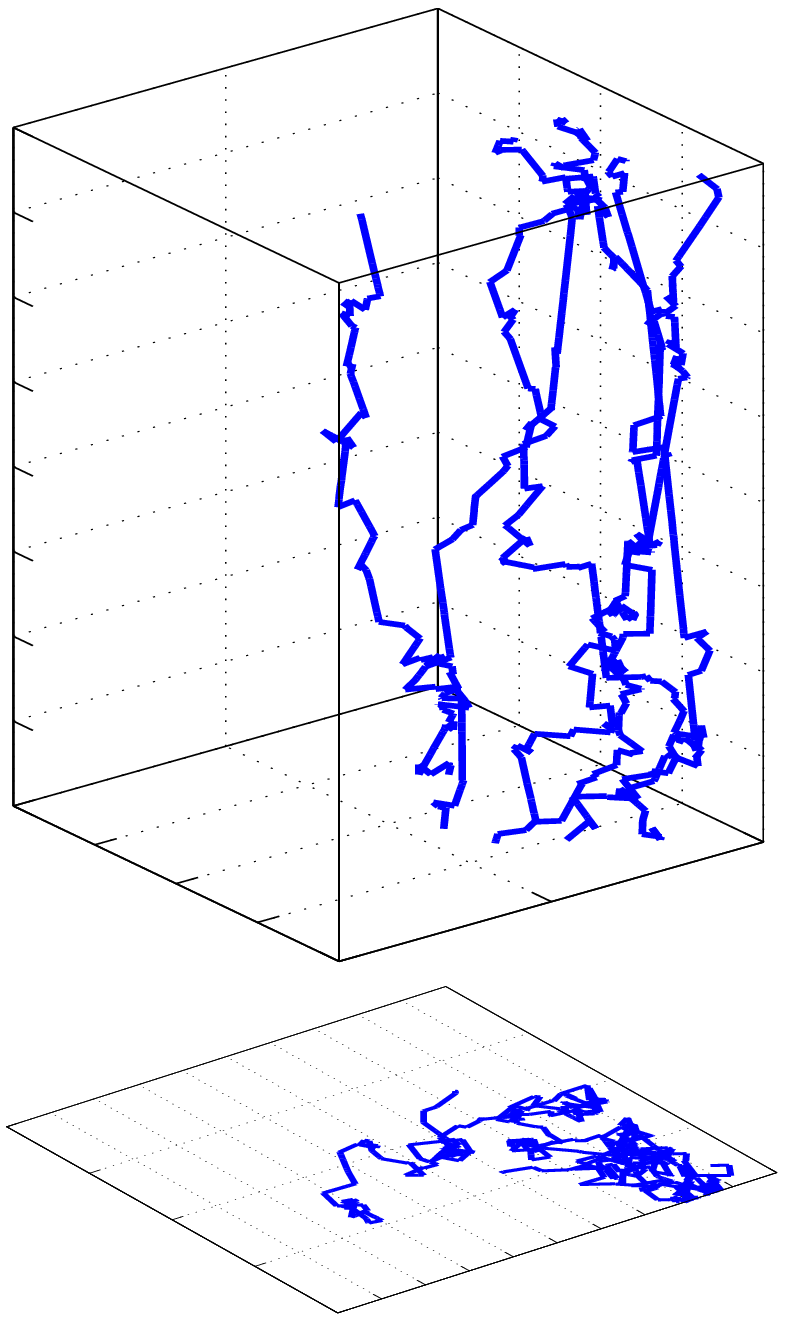}
      }
    }
    \caption{\label{ent_def} (a) Definition of the winding angle of two
      flux lines: For each z-coordinate the vector connecting the two lines
      is projected onto that basal plane. $z=0$ gives
      the reference line with respect to which the consecutive vectors for
      increasing $z$-coordinate have an angle $\phi(z)$. Once $\phi(z)>2\pi$
      the two lines are said to be entangled. (b) Top : A configuration of
      six entangled lines with splay disorder. Bottom: The projection of the
      line configuration on the basal plane, defining a connected
      cluster. (c) Six entangled lines with point disorder.}
  \end{center}
\end{figure}

In 3d lines can wind around each other
resulting in topologically non-trivial configurations which we analyze
here by computing the winding angle of all line pairs as indicated in
Fig.\ \ref{ent_def} (c.f.\ \cite{Drossel}). We define two lines to be
{\it entangled} when their winding angle becomes larger than
$2\pi$ \cite{entgl}. This provides a measure for topological entanglement
\cite{Samokhin}, since a topologically entangled pair 
can not be separated by any linear transformation in
the basal plane (i.e. the lines almost always would cut each other, if
one were shifted).

Sets or {\it bundles} of pairwise entangled lines are defined such
that a line belongs to a bundle if it is entangled at least with one
other line in the set. In the case of point disorder such bundles are
spaghetti-like --- i.e. topologically complicated and knotted sets of
one-dimensional objects which grow with increasing system height
leading finally to one giant bundle. In \cite{entgl} it was shown that
for point disorder there is a sharp transition from an non-entangled
phase without giant bundle to an entangled phase: The probability
$P_{perc}$ for having an entangled bundle of lines that spans the
system in the transverse direction jumps from 0 to 1 at a critical
height $H_c$ in the infinite system size limit ($L\to\infty$), 
and is described by the following finite size scaling form
\begin{equation}
P_{perc}=p(L^{1/\nu}(H-H_c))\;.
\label{eq:perc}
\end{equation}
One implication of this scaling form is that the location of the 
jump of $P_{perc}$ from 0 to 1 for finite size $L$ saturates at
some value $H_c$ in the limit $L\to\infty$, another that the 
jump width decreases with $L$. 

\begin{figure}\centering
  \subfigure[]{
    \includegraphics[angle=-90,width=0.99\columnwidth]{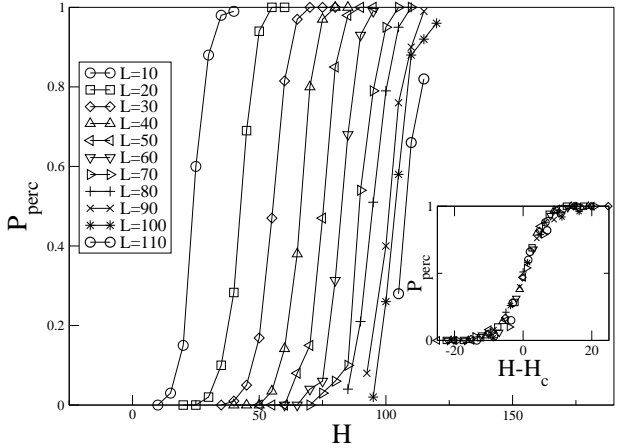}
  }
  \subfigure[]{
    \includegraphics[angle=-90,width=0.99\columnwidth]{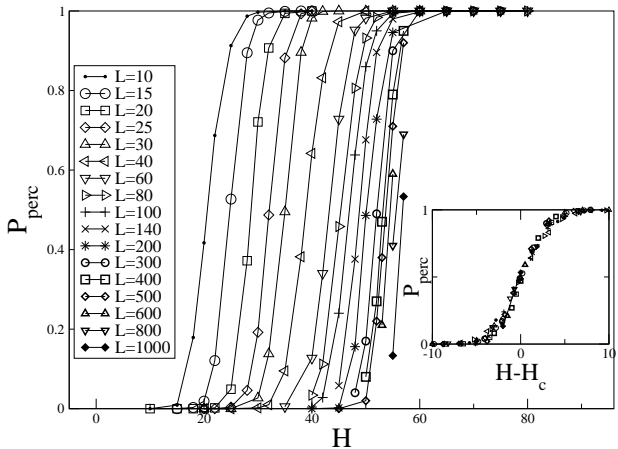}
  } 
  \caption{\label{perc} 
    Percolation probability as a function of the height of the system
    $H$: (a) splay disorder and (b) ballistic random walk.
    Insets: data shifted with $H_c$.
    For both the transition width does not change, there is only a shift of
    the jump $H_c(L)$.
}
\end{figure}

This appears not to be the case for splay disorder. As one can see
from Fig.\ \ref{perc}(a) $H_c$ does not saturate for the
computationally accessible system sizes ($L \lesssim 120$). From the
inset one can also see that the width of the transition does not
decrease.

For a comparison, we use
the following model consisting of random walks embedded into a box of size
$L\times L\times H$. Since the lines typically change their
direction only in the vicinity of the boundaries we consider ballistic
random walks. They start from a random point at the basal
plane and evolve towards the top level with a random tilt angle taken
from the same distribution as for the splayed columns. When the 
random walk meets the system boundary it is assigned a new tilt
angle. This is repeated until the path hits the top of the system.
This model corresponds to the high density limit where all
defects are occupied, i.e. $B=B_{\phi}$.

\begin{figure}\centering
  \includegraphics[angle=-90,width=0.99\columnwidth]{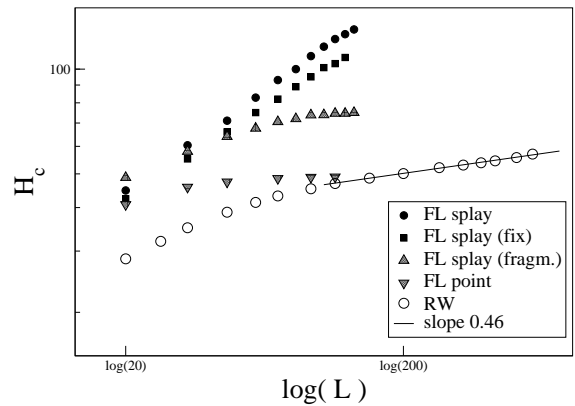}
  \caption{\label{hcvsL} A loglog plot of estimated critical system
    heights $H_c$ vs. logarithm of the system width $L$. From top:
    flux lines (FL) with free and fixed  
    (fix) starting points in splay disorder; flux lines in splay
    disorder with fragmentation (fragm.) and in point disorder;
    ballistic random walks (RW), solid line is a fit to few last data
    points.}
\end{figure}

We calculate the mutual winding angles  and check for entanglement
as before.
By comparing Fig.\ \ref{perc}(a) and Fig.\ \ref{perc}(b)
one observes no qualitative differences in the dependence of the
percolation probability $P_{perc}$ from the system height $H$.
No saturation is found for percolation height $H_c$ as
the system width is increased, though the growth is extremely
slow. According to our numerical results the transition height 
appears to grow 
asymptotically like $H_c\sim(\log L)^a$ with $a$ close to 0.5
(Fig. \ref{hcvsL}).  

\begin{figure}\centering
  \includegraphics[angle=-90,width=0.99\columnwidth]{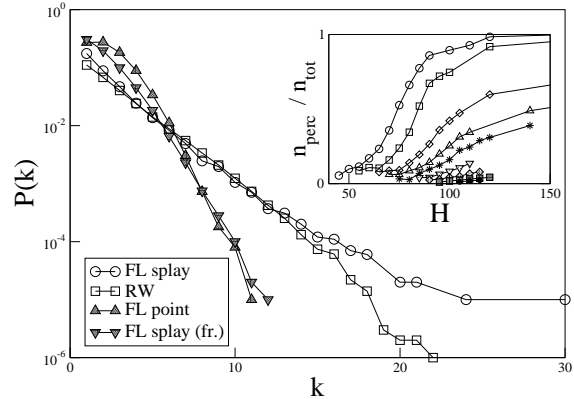}
  \caption{\label{entK} A plot with logarithmic $y$-axis of
    distribution of the degree of lines' entanglement $P(k)$. $k$ is
    defined as the  number of lines with which a given line is
    entangled. The following data sets are shown:
    flux lines in splay disorder with L=100, H=115 (ovals);
    ballistic random walks with L=100, H=39 (squares);
    flux lines in point disorder with L=100, H=48 (triangles up);
    flux lines in splay disorder with fragmentation L=100, H=70 (triangles
    down). The system heights are chosen so that the system is in the
    vicinity of the entanglement transition. Inset: strength of the
    percolating cluster of flux lines in splay disorder is plotted in
    a linear scale vs. system height $H$ for different system widths, L=30 
    (ovals) ... 110 (filled ovals).}
\end{figure}

The inset of Fig.\ \ref{entK} shows the relative size of the
percolating bundle $n_{perc}/n_{tot}$ for the data of Fig.\
\ref{perc}(a). It tends to zero for increasing system size also
for $H>H_c$. This is in contrast to what happens for conventional
percolation where $n_{perc}/n_{tot}=0$ exactly at the threshold and
then increasing as $\sim (H-H_c)^{\beta}$ with the critical exponent
$\beta$. 
Fig.\ \ref{entK} shows the distribution $P(k)$ of the number of lines
$k$ with which a given line is entangled (i.e. wind around it by an
angle larger than $2\pi$). One sees that the probability distribution
$P(k)$ is almost identical in the cases of splay disorder and
ballistic random walks, whereas $P(k)$ for point disorder is much
narrower. 

We conclude that in the case of splayed columnar disorder
the spanning bundles are formed mostly by few
lines which bounce from the boundaries.  Thus, there is 
no entanglement percolation transition in the case of splay 
disorder and no giant entangled cluster in the thermodynamic limit.

\begin{figure}\centering
    \includegraphics[angle=-90,width=0.99\columnwidth]{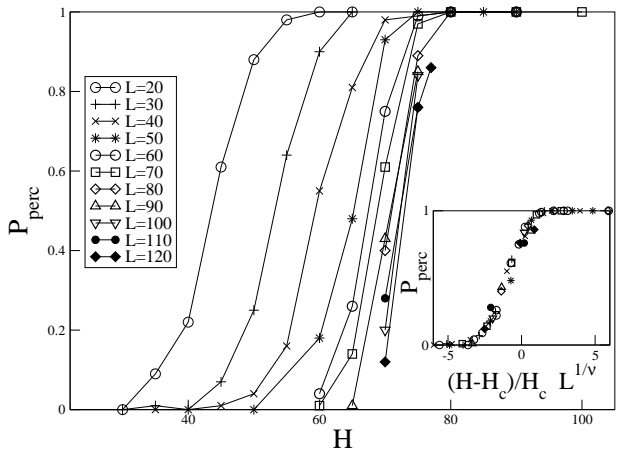}
  \caption{\label{fraperc} 
    Percolation probability as a function of the height of the system
    $H$ for 
    splay disorder with fragmentation. Inset: corresponding scaling
    plots, Eq. (\ref{eq:perc}). 
    Fragmentation causes the sharpening of the transition width
    with growing $L$ and data scales with the 2d percolation
    correlation length exponent $\nu=4/3$.
}
\end{figure}

This behavior changes when the splay
disorder is perturbed with attractive or repulsive point defects. 
Here we discuss results for relatively weak attraction.  The energy costs 
per unit length for a fraction $f$ of the bonds are now set to $u=0.5$ whereas
the rest of the bonds are as before.
Fig.\ \ref{fraperc} shows the behavior of the  percolation
probability corresponding to $f=0.1$. It indicates now the presence
of an entanglement transition following the conventional percolation
scenario (Eq. (\ref{eq:perc})). The transition height $H_c$ saturates 
in the limit 
$L\to\infty$ (Fig.\ \ref{hcvsL}) and the data obey finite size
scaling with the correlation length exponent $\nu=4/3$ of the 2d
percolation problem. 

The fraction $f$ tunes the typical length that a given line stays on
one defect. When this length exceeds the one needed to traverse the
system in the lateral direction one observes the properties of
pure splay disorder. Consequently as $f$ is decreased one needs
increasingly larger system sizes in order to observe the percolation
transition. The strength and the nature -- repulsive or attractive --
of the perturbing point disorder only changes the crossover size of 
the system as long as the perturbation is strong enough to induces
lines to change the defects. However, due to computational limitations
we did not attempt to find whether there is a threshold value for
strength of point disorder below which the lines cannot be entangled
also in the infinite system size limit.

\section{Summary and discussion}\label{sec:summ}

The ground state of a multi-line system with splay disorder 
has four different phases as a function of the system height
$H$. For small values, there is first a cross-over from trivial
behavior to the
single-line ground state roughening as lines start to jump
between columnar defects. For intermediate $H$, one observes
the collective regime where $\zeta$ depends on dimension
but not on the splay energy distribution. This arises when
the lines transverse wandering becomes of the same order of
magnitude as the average line distance.
Finally, there
is the cross-over to saturation for $L$ finite. The 
exponent values are a random walk-like $\zeta_{2d}=1/2$ and
a ballistic-like $\zeta_{3d}=1$. The cross-over between
the collective and single-line scalings is visible
in our 2d data. In 3d, this is not possible for numerical
restrictions. Note the fact that the 3d single line exponent
(which varies with the energy distribution) is {\em smaller}
than the collective one.

We have also considered the separability and entanglement, to
look at the other aspects of these systems as $H$ is varied.
The former measures the collective aspects of the line configuration,
which are most pronounced for $z$ small. In contrast to the
3d point disorder case, the winding of lines around
each other is suppressed, leading to the absence of entanglement in the
thermodynamic limit. In the context of our current model, this means
that for low line densities ($B<B_\phi$) 
the enhanced pinning expressed by increased critical current, is not
a collective effect.  

The results for the splay disorder are not expected to change
qualitatively when the flux lines are allowed to enter the homogenous bulk
between the defects. As discussed in end of Sec.\ \ref{sec:model}
in this case 
there will be a fraction of lines staying in the bulk throughout the
sample. Since these lines have no transverse fluctuations they 
reduce the average roughness by a constant factor.

Our numerical results suggest that inserting point-like defects
into splay disorder may even further enhance the flux line pinning 
as the entanglement transition of flux lines is recovered.
For large line densities ($B>B_\phi$) --
where experiments \cite{Kwok} imply entanglement -- one
could study the role of additional point disorder.
This brings new complications, already for single lines.
In the case of one attractive columnar defect competing with point
disorder in 2d one finds only a localized line,
whereas in 3d there a localization-delocalization
transition \cite{Balents}. Another study has found for
a mixture of many columnar defects and point disorder 
sub-ballistic behavior \cite{Arsenin}. 

Lidmar et al.  found by mixing in point disorder that in 2d splay
disorder always dominates on the long length scales whereas in 3d it
seems that strong enough fragmentation leads to point disorder
behavior \cite{Lidmar}.  Here we briefly present a few possible
scenarios associated with tuning the strength of point disorder with
fixed strength of splay disorder: (i) At large enough system sizes
point disorder will always dominate.  (ii) There is a crossover from
pure splay disorder behavior (no entanglement) to point disorder
behavior.  (iii) There is a third regime where lines take an advantage
of both splay and point disorder leading possibly to more efficient
entanglement than in the case of pure point disorder.  This would be
due to the large displacements along splayed defects. One can expect
that the response of the system to point disorder perturbations
depends on the line density.  It would be interesting to find the full
phase diagram including the optimal parameters from the point of view
of entanglement.

\acknowledgments
The Center of Excellence program of the Academy of Finland is\
thanked for financial support.

\end{document}